\documentclass[FBSedit,FBSmath,ecsub]{FBSsuppl}
\usepackage{amsfonts}
\usepackage{amssymb}
\usepackage{epsfig}
\title{Few-Body Resonances in Light Nuclei}
\author{Attila Cs\'ot\'o\thanks{\textit{E-mail address:} csoto@matrix.elte.hu}}
\institute{Department of Atomic Physics, E\"otv\"os University \\
P\'azm\'any P\'eter s\'et\'any 1/A, H--1117 Budapest, Hungary} 

\begin{document}

\maketitle
\begin{abstract}
We have localized several few-body resonances in light nuclei, using methods
which can properly handle two- or three-body resonant states. Among other 
results, we predict the existence of a three-neutron resonance, small 
spin-orbit splittings between the low-lying states in $^5$He and $^5$Li, the 
nonexistence of the soft dipole resonance in $^6$He, new $1^+$ states in 
$^8$Li and $^8$B, and the presence of a nonlinear amplification phenomenon 
in the $0^+_2$ state of $^{12}$C.
\end{abstract}

\section{Introduction}

In the past few years our theoretical understanding of the structure and
reactions of light nuclei has been greatly improved, thanks to new and powerful
methods and to the tremendous advances in computing power. Currently it is
possible to solve the bound-state problems of $A\leq 8$ nuclei, using realistic
two-body and three-body nucleon-nucleon (N-N) interactions, in a numerically
exact way \cite{GFMC}. The scattering problem is more difficult to deal with.
So far only the $A=3$ systems can be treated with the same high precision as
the bound states \cite{Glockle}. However, most of the states in light
nuclei are unbound resonances. As the most elaborate models cannot treat these
systems correctly for the time being, one can describe them either by using
methods which are unphysical at some level, or by treating the most important
degrees of freedom properly. We present here some of our recent results
achieved by following the second strategy
[3--12]. We concentrate mainly
on the physics motivations and the most interesting results. Further details
can be found in the original papers.

\section{Model}

We use a microscopic cluster model (RGM) description of nuclei, and
apply this model to systems whose wave functions contain two- or
three-cluster configurations with large weight. This ensures that those few
degrees of freedom (the one or two relative motions between the clusters) which
can be treated properly, are really the most important properties in the
problems. The wave functions of the two- and three-cluster systems look like
\begin{equation}
\Psi=\sum_{L,S}{\cal A} \Bigg \{ \bigg [ \Big [\Phi^{A}
\Phi^{B} \Big ]_S\chi_L(\mbox{\boldmath $\rho$})
\bigg ]_{JM} \Bigg\}
\label{wfn2}
\end{equation}
and
\begin{equation}
\Psi=\sum_{l_1,l_2,L,S}{\cal A} \Bigg \{ \bigg [
\Big [\Phi^{A} \Phi^{B} \Phi^{C}\Big ]_S
\chi_{[l_1,l_2]L}(\mbox{\boldmath $\rho$}_1,
\mbox{\boldmath $\rho$}_2)\bigg ]_{JM} \Bigg\},
\label{wfn3}
\end{equation}
respectively. Here ${\cal A}$ is the intercluster antisymmetrizer,
the $\Phi$ cluster internal states are translationally invariant
$0s$ harmonic-oscillator shell-model states, the $\mbox{\boldmath
$\rho$}$ vectors are the intercluster relative coordinates,
$l_1$ and $l_2$ are the angular momenta of the two relative
motions, $L$ is the total orbital angular momentum, $S$ is the
total intrinsic spin, and $[\ ]$ denotes angular momentum coupling. In
the case of three-cluster dynamics, all possible sets of relative
coordinates [$A(BC)$, $C(AB)$, $B(AC)$] and angular momentum
couplings are included in (\ref{wfn3}).

Putting (\ref{wfn2}) or (\ref{wfn3}) into the $N$-body
Schr\"odinger equation, we get equations for the unknown relative
motion functions $\chi$. For two-body (three-body) bound states
they are expanded in terms of (products of) Gaussian functions,
and the expansion coefficients are determined from a variational
principle for the energy. For two-body scattering states the $\chi$
functions are expanded in terms of Gaussian functions matched with
the correct asymptotics, and the expansion coefficients are
determined from the Kohn-Hulth\'en variational method for the $S$
matrix \cite{Kamimura}.

In scattering theory resonances are defined as complex-energy
solutions of the Schr\"odin\-ger equation that correspond to the
poles of the $S$ matrix (or equivalently the zeros of the Fredholm
determinant or Jost function). In order to obtain these complex
solutions, we implemented a direct analytic continuation of the $S$
matrix for two-cluster systems \cite{He5,pole}, and the complex scaling
method for three-cluster systems \cite{3br}.

For two-cluster systems we solve the Schr\"odinger equation for
the relative motion at complex energies with the 
boundary condition ($\rho\rightarrow \infty$)
\begin{equation}
\chi(\varepsilon,\rho)
\rightarrow H^-(k\rho)-\tilde S(\varepsilon) H^+(k\rho).
\end{equation}
Here $\varepsilon$ and $k$ are the {\it complex} energies
and wave numbers of the relative motions, and $H^-$ and
$H^+$ are the incoming and outgoing Coulomb functions,
respectively. The function $\tilde S$ has no physical
meaning, except if it is singular at the energy
$\varepsilon$. Then $\tilde S$ coincides with the physical
$S$ matrix, describing a purely outgoing solution, that is a
resonance. So we search for the poles of $\tilde S$ at complex
energies and extract the resonance parameters from
$\varepsilon=E_{\rm r}-i\Gamma/2$.

For the three-cluster systems we solve the eigenvalue problem of a
new Hamiltonian defined by
\begin{equation}
\widehat{H}_\theta=\widehat{U}(\theta)\widehat{H}
\widehat{U}^{-1}(\theta),
\end{equation}
where $\widehat{H}$ is the original many-body Hamiltonian and
$\widehat{U}$ is the complex scaling transformation which acts on a
function $f({\bf r})$ as $\widehat{U}(\theta)f({\bf r})=e^{3 i
\theta /2}f({\bf r}e^{i\theta})$.
In the case of a multicluster system the transformation
is performed on each dynamical coordinate (relative motion). The
solution of the complex-scaled Schr\"odinger equation
results in a spectrum with continuum cuts rotated by
$2\theta$ relative to the real energy axis, plus possibly a few
isolated complex points at the resonant and bound state poles
\cite{3br}.

\section{Resonances in ${\bf A=2-12}$ Nuclei}

We have used our model to study several selected resonances in $d(=p+n)$,
$^3n(=n+n+n)$, $^3p(=p+p+p)$, $^3{\rm H}(=p+n+n)$, $^3{\rm He}(=p+p+n)$,
$^4{\rm He}(=\{t+p,h+n\})$, $^5{\rm He}(=\alpha+n)$, $^5{\rm Li}(=\alpha+p)$,
$^6{\rm He}(=\alpha+n+n)$, $^6{\rm Li}(=\alpha+p+n)$, $^6{\rm
Be}(=\alpha+p+p)$, $^8{\rm Li}(=\alpha+t+n)$, $^8{\rm B}(=\alpha+h+p)$, and
$^{12}{\rm C}(=\alpha+\alpha+\alpha)$ \cite{fb}. Here $\alpha={^4{\rm He}}$, 
$t={^3{\rm H}}$, and $h={^3{\rm He}}$, and the cluster structures assumed in 
the model, are indicated. In most cases we used the Minnesota (MN) effective 
N-N interaction \cite{MN}, which gives a reasonably good overall description 
of the low-energy $N+N$ scattering and the bulk properties of the $^3$H, 
$^3$He, and $^4$He clusters. In certain cases the Eikemeier-Hackenbroich (EH), 
modified Hasegawa-Nagata (MHN), or Volkov (V1 and V2) forces \cite{forces} 
were applied.

\begin{figure}[!t]
\centering
\epsfig{file=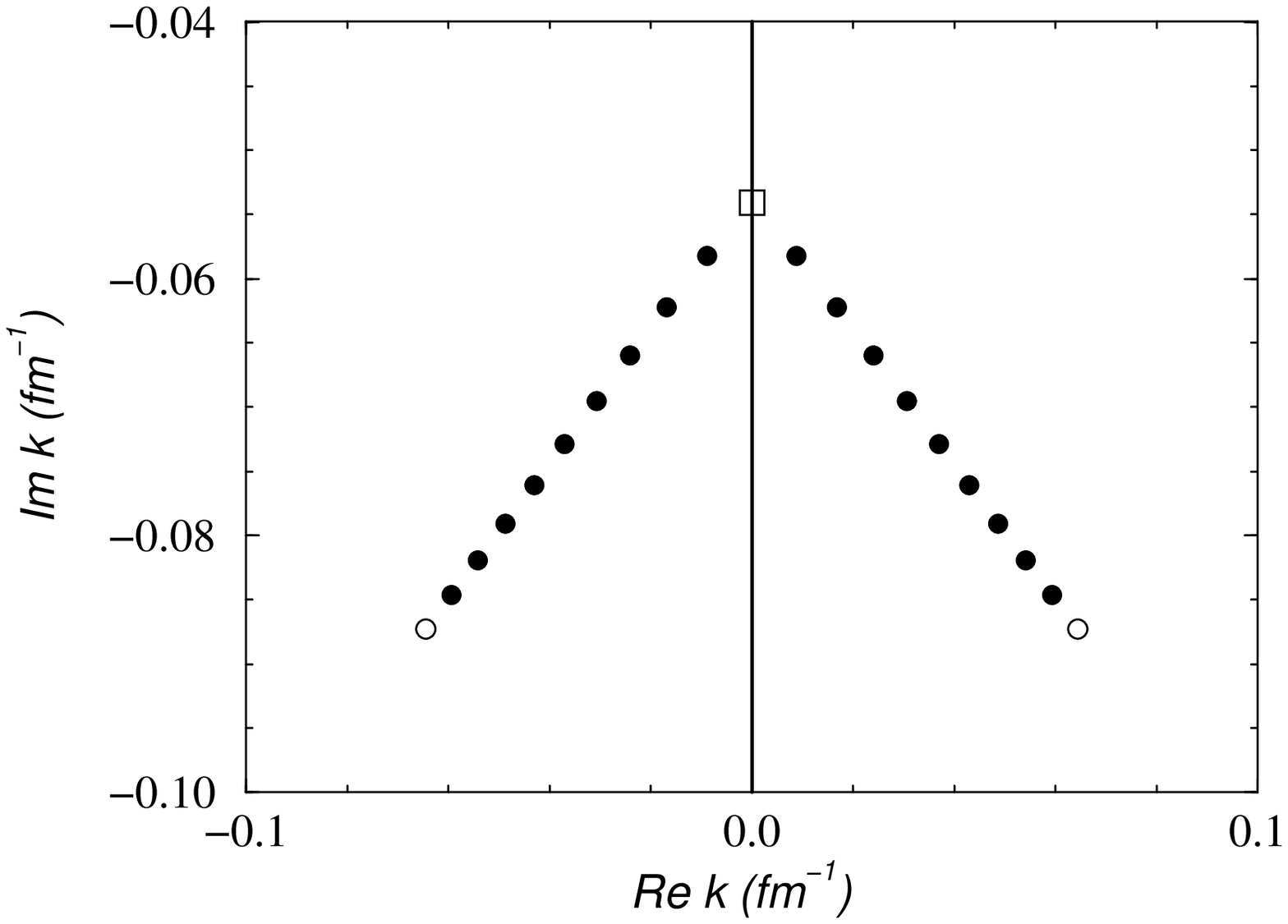,width=7.5cm}
\caption[]{Trajectories of the $^1S_0$ $N+N$ $S$-matrix poles.
The open square corresponds to the $n+n$ and $n+p$ poles,
wile the open circles denote the pair of conjugate poles in
the $p+p$ system. The filled circles come from calculations
where $c\cdot V_{\rm Coul}^{pp}$ is added to the $n+n$
interaction ($0<c<1$).}
\label{fig1}
\end{figure}

{\it A=2}: Our description of the virtual states in $^3$H and $^3$He (see
below) requires the good reproduction of the virtual states of the $N+N$ 
systems in the $^1S_0$ channel. We localized these states on the complex-energy
plane for the EH force by using the analytic continuation method \cite{H3}. In 
order to see how the differences between a neutral ($n+n$) and a charged 
($p+p$) two-body virtual state develop, we first localized the $n+n$ pole and  
then smoothly switched on the Coulomb force. As one can see in Fig.\ 
\ref{fig1}, the presence of the Coulomb force creates two poles from the 
single virtual state present in $n+n$, and moves them into the complex plane.
We note that in Ref.\ \cite{H3} our variational basis was not sufficiently
converged, leading to slightly incorrect pole positions. We correct this error
in Fig.\ \ref{fig1}. Our model gives the pole energies of the $n+n$ and $p+p$
states as $E_{nn}=-0.121$ MeV and $E_{pp}=(-0.143\pm i0.466)$ MeV, in excellent
agreement with the phenomenological values, $E_{nn}=-0.123$ MeV and
$E_{pp}=(-0.140\pm i0.467)$ MeV \cite{Kok}. Our interaction is charge
independent, leading to $E_{nn}=E_{np}$, therefore it is unable to reproduce
the phenomenological $E_{np}=-0.066$ MeV value \cite{Kok}. We would like to
emphasize that, as Fig.\ \ref{fig1} nicely demonstrates it, a virtual state
with pure imaginary wave number can exist only in a neutral $s$-wave two-body
system. The presence of a Coulomb-, centrifugal-, or three-body barrier does
not allow the appearance of an $S$-matrix pole on the negative imaginary $k$
axis.

{\it A=3}: The lightest nuclei where one could expect the existence of real
resonances with any experimental significance are the $A=3$ systems. The $3n$ 
and $3p$ nuclei are easier to handle, because the ${^3S_1}-{^3D_1}$ two-nucleon
channel is missing and there is no bound two-body subsystem present. We
searched for three-body resonances in the various partial waves using the MN
interaction, and found a $J^\pi=3/2^+$ resonance in $3n$ with $E_r=14$ MeV and 
$\Gamma=13$ MeV parameters, while the mirror $3p$ system has $E_r=15$ MeV and 
$\Gamma=14$ MeV \cite{3n}. The EH interaction gives somewhat smaller resonance 
energies. We should mention that some recent experiments did not see any 
evidence of these structures \cite{Palarczyk}. Thus, it would be highly 
desirable to repeat our calculations using fully realistic forces. The first 
step in this direction has been made in Ref.\ \cite{Witala}. So far those 
calculations could not be extended to the physical interactions, but the 
pole trajectories show that it is really the $3/2^+$ partial wave where one 
can expect a resonance, lying not very far from the real energy axis. 

In $^3$H and $^3$He there are evidences for the existence of $1/2^+$ virtual
states, both theoretically \cite{Moller} and experimentally
\cite{Brune}. However, to our knowledge, all existing calculations so far were
restricted to a simple configuration with a $^1S_0$ dinucleon plus the third
nucleon. We extended those works by taking into account some other important
channels, most importantly the $d+N$ configuration with a ${^3S_1}-{^3D_1}$
deuteron. Our model, which works only below the three-body breakup threshold,
gives $E_V=-1.62$ MeV and $E_V=(-0.43\pm i0.56)$ MeV for the energies of the 
$1/2^+$ virtual states in $^3$H and $^3$He, respectively, using the EH force
\cite{3n}. As the ground states of these nuclei are underbound in our model 
(due probably entirely to the lack of a three-body force), so are probably 
these virtual states. This underbinding leads to a too large $\vert E_V\vert$ 
for $^3$H, and to $^3$He states which lie too far from the imaginary $k$ axis.

{\it A=4}: $^4$He is the lightest nucleus with a well-established system of
resonances. The first excited state, $0^+_2$, which lies between the $^3{\rm
H}+p$ and $^3{\rm He}+n$ thresholds, is perhaps one of the most difficult
resonances to localize in $^4$He. Those approaches which cannot use correct
boundary conditions in their wave functions find this state several MeV above
the 3+1 thresholds \cite{Navratil}. Our model reproduces the phenomenological
$^1S_0$ phase shift in $^3{\rm H}+p$, which is the most important quantity
related to the $0^+_2$ resonance, rather well. We localize this state at
$E_r=93$ keV above the $^3{\rm H}+p$ threshold with $\Gamma=390$ keV width, as
a conventional resonance \cite{He4}. This result inspired us to repeat the
search for this state in the so-called extended $R$-matrix-model description of
the experimental data, because in the original search the resonance was not 
found \cite{A=4}. This time we could find this state in the extended 
$R$-matrix model with $E_r=114$ keV and $\Gamma=392$ keV parameters, which 
are close to the RGM values.

{\it A=5}: The $^5$He and $^5$Li nuclei offer one of the cleanest and easiest
testing grounds for resonance methods. Both of these systems have very strong
$\alpha+N$ clustering nature, and there exist precise $\alpha+N$ scattering
data. Still, there have been much controversy in the past few years concerning
the resonances of $^5$He and $^5$Li. Shell models, for example, suggest the
existence of low-lying positive parity states, most notably $1/2^+$ resonances
\cite{Navratil}. Furthermore, it appears that the phenomenologically extracted
spin-orbit splittings between the $3/2^-$ and $1/2^-$ states are so large that
their theoretical reproduction is hopeless \cite{Ajzenberg}. And, in general,
there seems to be big differences between the resonance parameters coming from
real-energy fits of certain reaction cross sections and those which are 
required by, e.g., the halo studies of $^6$He.

Using our RGM model, we localized the low-lying $^5$He and $^5$Li states as
poles of the scattering matrices \cite{He5}. Our calculations show that no
low-lying $1/2^+$ state of any experimental significance exists in these
nuclei. Extremely broad states ($\Gamma\gg 10$ MeV) can be found of course 
\cite{Comment}, as in almost any two-body system. However, they do not
have any observable effect on the $\alpha+N$ scattering. The calculated 
$3/2^-$ and $1/2^-$ 
resonance parameters are somewhat different from those coming from conventional
$R$-matrix analyses of the data \cite{Barker}, especially for the broader
resonances. However, if the $R$ matrices coming from the conventional data
fit, are extended to complex energies, then we get a good agreement with our
RGM results \cite{He5}. This finding emphasizes the fact that the correct
treatment of the asymptotics in the analysis of experimental data (e.g.\
through the extended $R$-matrix method) can substantially affect the results of
phenomenological analyses. Other calculations, using rather different models
and interactions, and treating the resonances properly, found resonance
parameters which are in excellent agreement with our results \cite{Aoyama}. 

We should mention that the prediction of the $1/2^+$ state and other positive
parity states by the shell model nicely demonstrates that the use of incorrect 
boundary conditions can lead to spurious states. In fact, if one could enlarge
the basis size in such models beyond any limit, then all their predicted states
(both the physical and spurious ones) would gradually move down to the lowest
breakup threshold. One could distinguish between the physical states and the
spurious ones by carefully analyzing their energy trajectories as functions of
the basis size. The physical resonances would show up as those which remain
stable for a relatively large interval of basis sizes \cite{He4,Comment,Hazi}. 

{\it A=6}: Our main motivation of studying the resonances of the $A=6$ nuclei
was to see if the predicted soft dipole state exists in the neutron-halo
nucleus $^6$He. It was suggested that the oscillation of the halo
neutrons against the $^4$He core in $^6$He would lead to a low-energy (a 
few MeV) dipole ($1^-$) resonance \cite{Hansen}. Break-up experiments performed
on $^6$He (and also on $^{11}$Li) really indicate low-lying bumps in the dipole
cross sections \cite{Janecke,Sacket}, which means that a concentration of the 
dipole strength is undoubtedly present in these systems. Its origin is,
however, questionable. Certain measurements show that these bumps come from a
direct break-up process and not from a long-lived dipole state \cite{Sacket}.
The main difficulty in interpreting the results comes from the fact that the
experiments can see only a one-dimensional projection (on the real-energy axis)
of a complex multisheeted energy surface, corresponding to the three-body
problem. Trying to find out from this one-dimensional image if the scattering
matrix has a $1^-$ (relative to the ground state) pole or not, is really
difficult. Theoretically the situation is much easier, although those methods
which are confined to real energies (e.g.\ the conventional shell model) face
the same difficulty as the experimental analyses. 

We searched for $\alpha+N+N$ resonances in $^6$He, $^6$Li, and $^6$Be by using
the complex scaling method in the cluster model \cite{He6}. We could find the
experimentally known states, but we did not see any evidence for the existence
of a $1^-$ state in $^6$He. This result has been confirmed by other
calculations using rather different methods and forces \cite{Kato}. We
should mention that those works, although do not see a $1^-$ state, indicate
the existence of several previously unknown resonances in $^6$He, 
like $0^+_2$ and $1^+$.
We believe that these new states are real, although they do not show up in our
model. It is possible that in our microscopic model these states would all be
rather broad, which could explain why our method, which becomes unstable for
broad states, cannot see them.

\begin{figure}[!t]
\centering
\epsfig{file=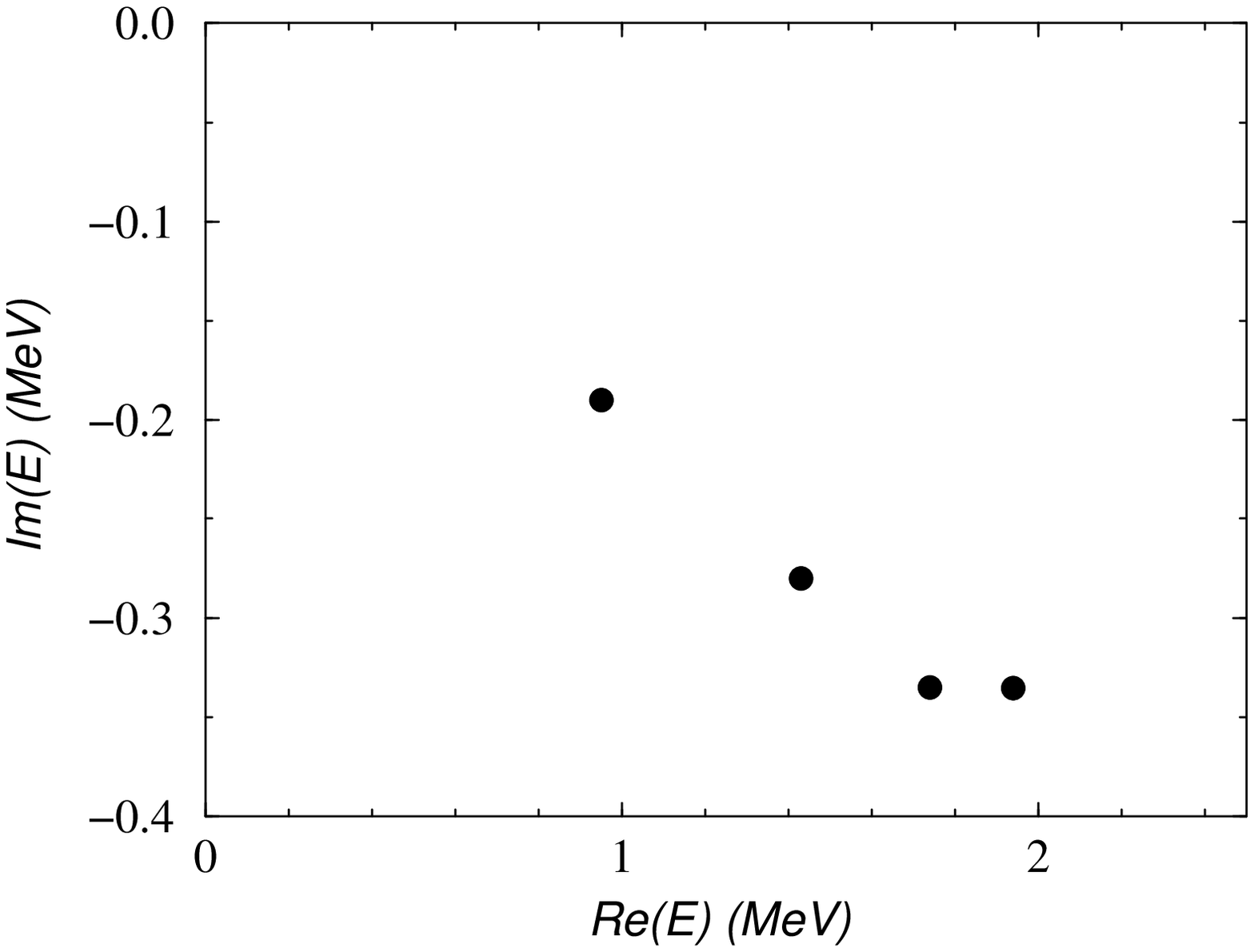,width=7.1cm}
\caption{Positions of the the first four poles of the $J^\pi=1^-$ $\alpha+n+n$ 
$S$ matrix of Ref.\ \protect\cite{Cobis} on the complex-energy plane.}
\label{fig2}
\end{figure}

Recently, several rather narrow $\alpha+n+n$ resonances of $^6$He were reported
in each partial wave in Ref.\ \cite{Cobis}. 
This finding contradicts all previous works
which, if they found any new state at all, indicated only a few rather broad
new resonances. We believe that the Ref.\ \cite{Cobis} results are wrong and
should be seen as a warning sign that the localization of resonances through
the $S$-matrix poles, although a very powerful method, can generate false
results if it is not done properly \cite{He6cont}. In Fig.\ \ref{fig2} we show
the complex-energy positions of the first four $1^-$ poles found in the Ref.\
\cite{Cobis} work (note that the first paper of Ref.\ \cite{Cobis} lists only
the first two poles in each partial waves; the others can be found in the 
second one). 
The distribution of these poles is clearly unphysical.

{\it A=8}: The $^7{\rm Li}(n,\gamma){^8{\rm Li}}$ and  $^7{\rm
Be}(p,\gamma){^8{\rm B}}$ reactions play important roles in astrophysics. The
first process takes place in certain inhomogeneous big-bang nucleosynthesis
models \cite{Rolfs}, while the second one makes $^8$B in the sun, which
produces the highest-energy solar neutrinos with substantial flux
\cite{Bahcall}. Although astrophysically only the very low-energy (practically
$E=0$) cross sections are important, these values can be substantially
influenced by the higher-energy continuum structures present in $^8$Li and
$^8$B. For example, the extrapolation of the astrophysical $S$ factor $S(E)$   
of the $^7{\rm Be}(p,\gamma){^8{\rm B}}$ reaction could be affected by the 
existence of a second $1^+$ state at low energies \cite{offsh}. 
With this motivation, we searched for low-energy $1^+$ 
states in $^8$Li and $^8$B \cite{B8cont}. After tuning the MHN force to 
precisely reproduce the known parameters of the 0.632-MeV $1^+$ state of $^8$B 
(relative to $^7{\rm Be}+p$), we indeed found new $1^+$ states. In $^8$B it is
situated at $E_r=1.28$ MeV and has a $\Gamma=0.56$ MeV width, while in $^8$Li
it lies right at the $^7{\rm Li}+n$ threshold. We would like to emphasize that
so far there is no experimental evidence which would support our results.
Nevertheless, given the important consequences of such states if they
exist, we think that further experimental and theoretical studies of these
possible structures would be desirable. One can see, e.g., that the presence 
of such a resonance would really affect the extrapolation of the experimental 
$S(E)$ of the $^7{\rm Be}(p,\gamma){^8{\rm B}}$ reaction from higher energies 
down to $E=0$ \cite{B8cont}.

{\it A=12}: Recently, we studied the low-lying $3\alpha$ resonances of $^{12}$C
\cite{C12}. We were able to reproduce the known resonances, and we believe that
for the first time we showed that the $0^+_2$ state is a genuine three-alpha
resonance of $^{12}$C. This level plays an important role in astrophysics, as
virtually all the carbon in the Universe is synthesized through it
\cite{Rolfs}. But this $0^+_2$ resonance is interesting for another reason,
too. It possesses a rather curious feature, which we call nonlinear quantum
amplifying \cite{amplif}. If we change the strength of the N-N interaction by
0.1\%, then the resonance energy of this state, relative to the
$3\alpha$ threshold, changes a lot more, by almost 10\%. One can study this
response to small perturbations also in other nuclei. It turns out that the
relatively deeply bound states give a response which is comparable in size to
the perturbation. However, as one moves close to the edge of stability, the
effect of a small perturbation can get enormously amplified in the energy. This
effect is caused by the fact that the residual interactions between the
clusters, to which the nucleus breaks up, go toward zero much more mildly
than the binding or resonance energy itself, as we go toward the break-up 
point.

\begin{figure}[!t]
\centering
\epsfig{file=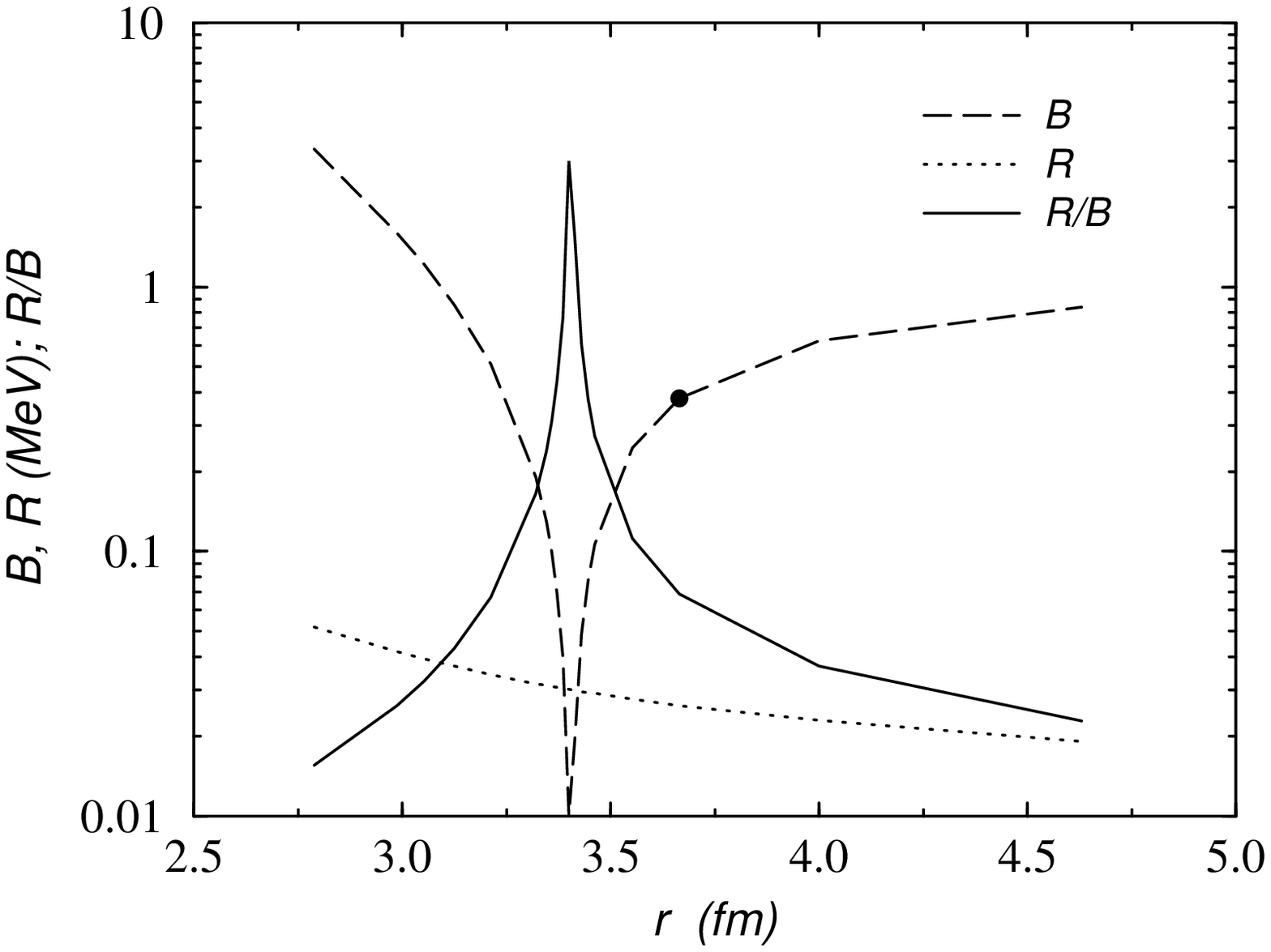,width=8.2cm}
\caption{The energy ($B=\vert E\vert$, where $E$ is the binding energy
or resonance energy, relative to the breakup threshold), the response
($R=\vert E_p-E_{p\times 1.001}\vert$, where $E_p$ and $E_{p\times
1.001}$ are the binding energies or resonance energies corresponding
to a given N-N force and another one which is stronger by 0.1\%,
respectively), and the $R/B$ ratio calculated for several artificial
$0^+_2$ states of $^{12}$C, as functions of the radius of the state. The 
N-N interaction is chosen to be the MN force in each case, with the strengths 
multiplied by a number $p$ (see the text). The black dot shows the real 
physical $0^+_2$ state, given by our model.}
\label{fig3}
\end{figure}

This behavior is demonstrated in the case of the $0^+_2$ state of $^{12}$C in
Fig.\ \ref{fig3}.
We generated several artificial $0^+_2$ states by changing the strength of the
N-N force (multiplying all strength of the strong force by a number $p$). Then,
for each artificial state (represented by its binding/resonance energy and its
radius) we calculated the response, that is the change of the binding/resonance
energy caused by a 0.1\% increase in the N-N strength. As one can see in Fig.\
\ref{fig3}, the response (which is closely related to the residual interaction)
really behaves very differently than the binding/resonance energy, as we
approach the break-up point. This naturally leads to the possibility of huge
amplifications. We believe that this phenomenon is a common feature of nuclei
lying at the edge of stability. 

The strong sensitivity of the resonance energy of the $0^+_2$ state of $^{12}$C
to the N-N force has a spectacular consequence in astrophysical carbon 
synthesis. Careful studies of all the details of the process show that a mere 
0.5\% change in the strength of the N-N force would lead to a Universe where 
virtually no carbon or oxygen exists \cite{carbon}. This makes 
carbon production one of the most fine-tuned processes in astrophysics, 
leading to interesting consequences for the possible values of some 
fundamental parameters of the Standard Model \cite{Jeltema}.

\section{Conclusions}

We have presented selected examples of some interesting few-body resonances in
light nuclei. We believe that the investigation of these and other resonance
structures, using methods which can properly handle them, offers a rich source
of information on many-body dynamics, nucleon-nucleon interaction,
shell-structure, etc.

\begin{acknowledge}
This work was supported by Grants from the OTKA Fund (D32513), the Education 
Ministry (FKFP-0242/2000--0147/2001) and the National Academy (BO/00520/98) of 
Hungary, and by the John Templeton Foundation (938-COS153). We are grateful 
to G.~M.\ Hale and H. Oberhummer for many useful discussions. 
\end{acknowledge}

\end{document}